\documentstyle[12pt]{article}

\makeatletter

\makeatletter

\if@twoside
\else
\fi
\advance\topmargin by -3mm

\ifcase \@ptsize
\or 
\or 
\fi

\def\@citex[#1]#2{%
\if@filesw \immediate \write \@auxout {\string \citation {#2}}\fi
\@tempcntb\m@ne \let\@h@ld\relax \def\@citea{}%
\@cite{%
 \@for \@citeb:=#2\do {%
 \@ifundefined {b@\@citeb}%
 {\@h@ld\@citea\@tempcntb\m@ne{\bf ?}%
  \@warning {Citation `\@citeb ' on page \thepage \space undefined}}%
 {\@tempcnta\@tempcntb \advance\@tempcnta\@ne%
 \@tempcntb\number\csname b@\@citeb \endcsname \relax%
 \ifnum\@tempcnta=\@tempcntb %
 \ifx\@h@ld\relax%
 \edef \@h@ld{\@citea\csname b@\@citeb\endcsname}%
 \else%
 \edef\@h@ld{\ifmmode{-}\else--\fi\csname b@\@citeb\endcsname}%
 \fi%
 \else
 \@h@ld\@citea\csname b@\@citeb \endcsname%
	\let\@h@ld\relax%
      \fi}%
    \def\@citea{,\penalty\@highpenalty\,}%
  }\@h@ld
}{#1}}
\@addtoreset{equation}{section}
\def\section{\@startsection {section}{1}{\z@}{-3.5ex plus -1ex minus
 -.2ex}{2.3ex plus .2ex}{\large\bf\centering}}
\def\subsection{\@startsection{subsection}{2}{\z@}{-3.25ex plus%
 -1ex minus -.2ex}{1.5ex plus .2ex}{\sc}}
\gdef\@publabel{\hfil}
\gdef\@pubdate{\null}
\gdef\@pubnumber{\null}
\gdef\@author{\null}
\gdef\@title{\null}
\gdef\@abstract{\null}
\long\def\pubdate#1{\gdef\@pubdate{#1}}
\long\def\pubnumber#1{\gdef\@pubnumber{#1}}
\long\def\publabel#1{\gdef\@publabel{#1}}
\long\def\author#1{\gdef\@author{#1}}
\long\def\title#1{\gdef\@title{#1}}
\long\def\abstract#1{\gdef\@abstract{#1}}
\def\titlerelax{
\let\maketitle\relax
\let\settitleparameters\relax
\let\consolidatetitle\relax
\let\inittitlepage\relax
\let\finishtitlepage\relax
\let\titlepagecontents\relax
\let\multithanks\relax
\let\titlebaselines\relax
\let\@makepub\relax
\let\@maketitle\relax
\let\@makeauthor\relax
\let\@makeabstract\relax
\let\@maketitlenote\relax
\let\thanks\relax
\let\titlerelax\relax}
\def\titleclean
{\gdef\@titlenote{}
\gdef\@abstract{}
\gdef\@author{}
\gdef\@title{}
\gdef\@pubdate{}\gdef\@pubnumber{}\gdef\@publabel{}
\gdef\@dpublabel{}
}

\def\@makepub{\vbox to \z@{\hbox to \textwidth{\hfill
\@publabel \hfill
\llap{\parbox[t]{0.33\textwidth}{\raggedleft\@pubnumber}}}%
\vss}}
\def\@maketitle{\vskip 60pt \begin{center}
 {\LARGE \@title \par}
 \end{center}}

\def\@makeauthor{{%
\def\and{\smallskip {\normalsize \rm and\smallskip }}
\def\And{\medskip {\normalsize \rm and\\}\medskip}
\long\def\address##1{{\def\and{\\and\\}\medskip
				{\small \it \\##1\\}
}}
{\centering
 \vskip 3em
 \large \lineskip .75em
 \@author}
 \par}}
\def\@makedate{\vskip 1.5em
 \rightline{\raggedright \small \noindent\@pubdate}}
\def\@makeabstract{\vskip 1.5em
{\small
\begin{center}
{\bf ABSTRACT\vspace{-.5em}\vspace{0pt}}
\end{center}
\quotation \@abstract \endquotation}}
\def\maketitle{\titlepage
\let\footnotesize\small \setcounter{page}{0}
\@makepub
\@makedate
\vfil
\@maketitle
\@makeauthor
\vfil
\@makeabstract
\@thanks
\vfil
\titlerelax \titleclean
\setcounter{footnote}{0}
}

\def\bigans{y }

\read-1 to\answ
\ifx\answ\bigans
\read-1 to\bnsw
	\ifx\bnsw\bigans 

\ifcase\@ptsize
 \font\tenmsa=msam10
 \font\sevenmsa=msam7
 \font\fivemsa=msam5
 \font\tenmsb=msbm10
 \font\sevenmsb=msbm7
 \font\fivemsb=msbm5
\or
 \font\tenmsa=msam10 scaled \magstephalf
 \font\sevenmsa=msam8
 \font\fivemsa=msam6
 \font\tenmsb=msbm10 scaled \magstephalf
 \font\sevenmsb=msbm8
 \font\fivemsb=msbm6
\or
 \font\tenmsa=msam10 scaled \magstep1
 \font\sevenmsa=msam8
 \font\fivemsa=msam6
 \font\tenmsb=msbm10 scaled \magstep1
 \font\sevenmsb=msbm8
 \font\fivemsb=msbm6
\fi

\newfam\msafam
\newfam\msbfam
\textfont\msafam=\tenmsa  \scriptfont\msafam=\sevenmsa
  \scriptscriptfont\msafam=\fivemsa
\textfont\msbfam=\tenmsb  \scriptfont\msbfam=\sevenmsb
  \scriptscriptfont\msbfam=\fivemsb

\def\hexnumber@#1{\ifnum#1<10 \number#1\else
 \ifnum#1=10 A\else\ifnum#1=11 B\else\ifnum#1=12 C\else
 \ifnum#1=13 D\else\ifnum#1=14 E\else\ifnum#1=15 F\fi\fi\fi\fi\fi\fi\fi}

\def\msa@{\hexnumber@\msafam}
\def\msb@{\hexnumber@\msbfam}

\def\Bbb{\ifmmode\let\next\Bbb@\else
 \def\next{\errmessage{Use \string\Bbb\space only in math mode}}\fi\next}
\def\Bbb@#1{{\Bbb@@{#1}}}
\def\Bbb@@#1{\fam\msbfam#1}
	\else
		\ifx\cnsw\bigans 

\ifcase\@ptsize
 \font\tenmsx=msxm10
 \font\sevenmsx=msxm7
 \font\fivemsx=msxm5
 \font\tenmsy=msym10
 \font\sevenmsy=msym7
 \font\fivemsy=msym5
\or
 \font\tenmsx=msxm10 scaled \magstephalf
 \font\sevenmsx=msxm8
 \font\fivemsx=msxm6
 \font\tenmsy=msym10 scaled \magstephalf
 \font\sevenmsy=msym8
 \font\fivemsy=msym6
\or
 \font\tenmsx=msxm10 scaled \magstep1
 \font\sevenmsx=msxm8
 \font\fivemsx=msxm6
 \font\tenmsy=msym10 scaled \magstep1
 \font\sevenmsy=msym8
 \font\fivemsy=msym6
\fi

\newfam\msxfam
\newfam\msyfam
\textfont\msxfam=\tenmsx  \scriptfont\msxfam=\sevenmsx
  \scriptscriptfont\msxfam=\fivemsx
\textfont\msyfam=\tenmsy  \scriptfont\msyfam=\sevenmsy
  \scriptscriptfont\msyfam=\fivemsy

\def\hexnumber@#1{\ifnum#1<10 \number#1\else
 \ifnum#1=10 A\else\ifnum#1=11 B\else\ifnum#1=12 C\else
 \ifnum#1=13 D\else\ifnum#1=14 E\else\ifnum#1=15 F\fi\fi\fi\fi\fi\fi\fi}

\def\msx@{\hexnumber@\msxfam}
\def\msy@{\hexnumber@\msyfam}

\def\Bbb{\ifmmode\let\next\Bbb@\else
 \def\next{\errmessage{Use \string\Bbb\space only in math mode}}\fi\next}
\def\Bbb@#1{{\Bbb@@{#1}}}
\def\Bbb@@#1{\fam\msyfam#1}

\else
\def\Bbb#1{{\bf #1}}
		\fi
	\fi
\else
\def\Bbb#1{{\bf #1}}
\fi

\def\thebibliography#1{\section*{References\@mkboth
 {REFERENCES}{REFERENCES}}\list
 {[\arabic{enumi}]}{\settowidth\labelwidth{[#1]}\leftmargin\labelwidth
 \advance\leftmargin\labelsep
 \usecounter{enumi}}
 \def\newblock{\hskip .11em plus .33em minus .07em
 \sloppy\clubpenalty4000\widowpenalty4000
 \itemsep=0pt
 \small
 \sfcode`\.=1000\relax}}

\makeatother

\def\choose#1#2{#1}

\def\a{\alpha}

\def\b{\beta}
\def\blank#1{}

\def\cdd{{\cdot}}
\def\cev#1{\langle #1 \vert}
\def\cL{{\cal L}}

\def\en{\end{equation}}
\def\eq{\begin{equation}}
\def\enn{\end{eqnarray}}
\def\eqq{\begin{eqnarray}}

\def\hf{\frac 12}

\def\mass{\mu}
\def\mno{{\textstyle {\circ\atop\circ}}}

\def\reff#1{(\ref{#1})}

\def\t{\tau}

\def\vec#1{\vert #1 \rangle}

\begin{document}

\pubnumber{DAMTP--94--27\\ hep-th/9404065}
\title{Quantum Mass corrections for $C_2^{(1)}$ Affine Toda theory
solitons}

\author{
G.\ M.\ T.\ Watts%
\thanks{E-mail address: \tt g.m.t.watts@damtp.cambridge.ac.uk}%
\address{St.\ John's College, St.\ John's Street, Cambridge, CB2 1TP,
U.\ K.
\\ and \\
DAMTP, University of Cambridge, Silver Street, Cambridge, CB3 9EW,
U.\ K.}%
}

\abstract{
We calculate the quantum mass corrections to the solitons in the
$C_2^{(1)}$ Affine Toda field theory following Hollowood. We find that
the ratio of the masses of the two solitons is not constant.
}

\pubdate{revised 20 April, 1994 }

\maketitle

\openup1\jot

\section{Introduction}

An affine Toda field theory is a theory of scalar fields in two
dimensions with exponential interactions. There is an affine Toda field
theory associated with each Kac--Moody algebra, with the interactions
given by the simple roots of the algebra. If we denote the fields by
$\phi^i$, and the simple roots of the Kac--Moody algebra by
$\alpha_a^i$, then the action takes the form
\eq
\cL
=
\hf \partial^\mu\phi\cdd\partial_\mu\phi
- 
\frac{\mass ^2}{\b^2} \sum_{a}
n_a
        \left[ 
        \exp( \beta\alpha_a\cdd\phi )
        -1 \right] 
\;,
\label{eq.lag}
\en
where $\b$ is the coupling constant, $\mass $ is the mass scale and $n_a$
are numbers chosen so that $\phi=0$ is the minimum of the potential.
What makes the Toda theories special is that
classically they are integrable with conserved quantities whose
spins are given by the exponents of the affine algebra.
The Kac--Moody algebras maybe be classified in a similar manner to
finite Lie algebras by Dynkin diagrams which encode the inner products
of the simple roots
(for details of Kac--Moody algebras and their classification see
e.g. \cite{Kac1}).

For a long time Affine Toda field theories were only studied
with the coupling constant $\b$ real. As Quantum field theories these
are theories of $n$ scalar particles, and the presence
of higher spin conserved quantities in the quantum theory implies
that the scattering preserves individual particle momenta, and that
the S--matrix factorises on the two particle scatterings (see \cite{AfTod}
for details).
\blank{
Classically the masses of the particles obey many nice relations, and
in the  quantum  theory  they are even more interesting.
The main result is that the masses are dependent on $\b$ and $\mass $, but
for the self--dual theories the ratio is invariant. For the dual
pairs, the masses of one theory flow from their classical values at
weak coupling to those of the dual theory at string coupling.
}
\blank{
There are problems in defining the theories for imaginary $\b$ (except
for the case $A_1^{(1)}$ which gives Sine-Gordon theory), but
Hollowood realised that there could be interesting results for these
theories. In particular they could be naturally related to perturbed
conformal field theories, as imaginary coupling constant conformal
Toda theories are naturally related to conformal field theories.
}

The imaginary coupling constant theories have a very different
spectrum, as can be seen from the Sine Gordon model. The potential is
periodic and so there are finite energy solutions which interpolate
different vacua which may be described in terms of interacting
solitons. There are also `breather' solutions which appear as bound
states of solitons.

For Affine Toda theories other than the Sine-Gordon model, there are
difficulties arising because the potential is not real. However,
Hollowood was able to construct sensible soliton solutions for the
$A_n^{(1)}$ theories  for which, although the energy density was not
real, the energy itself was \cite{Holl3}. This result was succeeded by
many others
on classical soliton spectra of imaginary coupling Affine Toda
theories \cite{MMcg1,CZhu1,CFGo1,ACFe1,ACFe2}, of which that of Olive
et al.\ is the most complete
\cite{OTUn1}. They
have found a complete set of soliton solutions for the untwisted
Affine Toda field theories, and the classical mass spectra of the
solitons. They found the remarkable result that the masses of the
solitons in the theory $g^{(1)}$ were those of the particles in the
theory based on $ \left( g^\vee \right)^{(1)}$, a result reminiscent
of, but not the same as that found for the quantum masses in real
coupling theories.

Hollowood then turned his attention to the quantum mass corrections to
the solitons in the $A_n^{(1)}$ theories \cite{Holl2}. He followed the
method of
Dashen, Hasslacher and Neveu \cite{DHNe1}, which is described in
the book by
Rajaraman \cite{Raja1}. He found that, as with the particle spectrum
in the
$A_n^{(1)}$ models, the ratios of the soliton masses were unchanged by
the quantum corrections.

It is obviously an interesting problem to find the quantum mass
corrections for the non-simply laced Affine Toda solitons.
In this letter we perform this calculation for the two solitons of
the $C_2^{(1)}$ theory.

\section{The $C_2^{(1)}$ Affine Toda field theory}

The $C_2^{(1)}$ Affine Toda field theory is a theory of 2 scalar
fields. The Lagrangian is given by eqn.\ (\ref{eq.lag})
with
\eq
\begin{array}{rclrcl}
n_1 &=& 2, & \a_1 &=& (  0, 1) \\
n_2 &=& 1, & \a_2 &=& (  1,-1)\\
n_0 &=& 1, & \a_0 &=& ( -1,-1)
\end{array}
\en
Expanding to second order in $\b$ we find the mass matrix
\eq
\phi \cdd M \cdd \phi = \mass ^2 ( 2 \phi_1^2 + 4 \phi_2^2)
\en
so that the `classical' masses of the particles are
 $\sqrt 2 \mass $ and $2 \mass $. We call these  particles type (i) and (ii)
respectively.

\section{The Soliton solutions of the $C_2^{(1)}$ theory}

We can write a soliton solution $\phi$ of the equations of motion
as
\eq
\phi = -\frac 1\b \sum \a_j^\vee  \ln \tau_j
\en
where $\a_j^\vee = 2 \a_j / (\a_j)^2 $.
The solitons of the $C_n^{(1)}$ theories may be obtained from the
soliton solutions to the $A_3^{(1)}$ Affine Toda theory. There are two
classes of single solitons of the $C_2^{(1)}$ theory which are derived
{}from a double soliton and a single soliton of the $A_3^{(1)}$ theory
respectively. If we denote these by type I and type II, they are given
by
\eq
\begin{array}{r|cc}
	& \hbox{type I} & \hbox{type II} \\[1mm]
\hline\\[-3mm]
\t_0    & 1 + 2 X + \hf X^2     & 1 + X \\[1mm]
\t_1    & 1       + \hf X^2     & 1 - X \\[1mm]
\t_2    & 1 - 2 X + \hf X^2     & 1 + X \\[1mm]
X       & \;\;

q \exp( \sqrt 2 \mass  ( x \cosh \eta + t \sinh \eta )  )
	  \;\;
	& \;\;
q \exp( 2 \mass  ( x \cosh \eta + t \sinh \eta ) )
	  \;\;
\end{array}
\en
In each of these solutions $q$ is a free parameter which fixes the
topological charge and position of the soliton and $\eta$ is the
rapidity of the soliton. As can be seen
these solitons are uniquely identified with a particle by the
exponent in the exponential fall off of the soliton at rest ($\eta =
0$). The type I soliton can be
thought of as a bound state of type (i) particles, and
the type II soliton as a bound state of type (ii) particles.
The type I solitons have topological charge in the 4 dimensional
representation of $C_2$, and type II in the 5 dimensional
representation.
Although the single and multiple solitons of $C_n^{(1)}$ have been
given elsewhere, here we have used \cite{OTUn1} to find all the
solutions given in this letter, although our normalisations may
differ.

The masses of these solitons can be deduced using the classical
formula for the Hamiltonian $H = \int \!d\!x T_{00}$ in terms of the
energy-momentum tensor,
\eq
T_{\mu \nu}
=
\partial_\mu \phi \frac{\partial \cL}{\partial (\partial^\nu \phi)}
- \eta_{\mu \nu} \cL
\en
giving the answers \cite{OTUn2}
\eq
H =
\int d\!x\,
\left(
\hf \dot\phi\dot\phi + \hf \phi'\phi'
+
\frac{\mass^2}{\b^2} \sum_a n_a
        \left[ 
        \exp( \b \alpha_a \cdd \phi )
        -1 \right] 
\right)
\label{eq.ham}
\en
\eq
M_I = -\frac{ 16 
\sqrt 2 \mass }{\b^2}
\;\;,\;\;\;\;
M_{II} = -\frac{ 16 
\mass }{\b^2}
\en


In \cite{DHNe1} Dashen et al.\
showed how to calculate quantum mass corrections to kink and soliton
solutions in 2 dimensional quantum field theory.
The standard approach requires the introduction of a set of harmonic
osscilators for each bounded solution to the linearised equations of
motion. These give an infinite contribution to the Hamiltonian from
the energies  of the oscillator vacuum states, and from the potential term as
well. Since the linearised
equations have different bounded solutions in the vacuum and in the
background of a soliton, this will lead to one source of mass
corrections. The result will be divergent,
but the divergences cancel when we examine the ratio of the soliton masses.
We set $\hbar = 1$ throughout this letter.

\subsection{Zero-point energy subtraction}

In the vacuum sector the linearised equations of motion are
\eq
\ddot{\phi} - \phi''
=
\left(
{\mass ^2} \sum_{a} n_a \alpha_a \alpha_a
\right)
\cdd\phi
\;,
\label{eq.l.eom.vac}
\en
\choose
{%
We put the theory in a box of length $L$ and impose $\phi$ periodic.
}{%
We put the theory in a box of length $L$ and impose $\phi=0 $ at the
edges of the box.
}
The solutions to (\ref{eq.l.eom.vac}) are
\eq
\xi_a \exp (i k x - i \omega t) +
\xi^*_a \exp (-i k x + i \omega t)
 \;,\;\; a=1,2
\en
where $a$ labels the two mass eigenstates,  $\omega^2 = k^2 + m^2$
and $ k = 2 n \pi / L $.
We shall want later to take the limit $L \to \infty$ in which case
the ground state energy is
the quadratically divergent integral
\eq
E_0^{(vac)} =
\lim_{\Lambda \to \infty} \sum_{a=1,2}
L \int_{-\Lambda}^\Lambda \frac{d\!k}{4\pi} \sqrt{ k^2 + m_a^2 }
\label{eq.e0}
\en
where we have introduced a momentum cutoff $\Lambda$.
In a soliton background we can find all the solutions to the
linearised equations from the two soliton solutions. Roughly speaking,
a two soliton solution will have two free parameters $q, q'$. Since
this is a solution for arbitrary $q'$, the coefficient of $q'$ must
satisfy the linearised equations of motion in the background to the
other soliton.
We give the three two--soliton solutions in table \ref{tab.2sol}.
\begin{table}
\[
\begin{array}{r|ccc}
	& \hbox{type I -- I}    & \hbox{type I -- II}
				& \hbox{type II -- II}
\\[1mm]
\hline\\[-3mm]
\t_0    &
\matrix{
1 + 2 X + \hf X^2
  + 2 Y + \hf Y^2 \cr
  + 2 XY (\xi + \zeta)
  + X Y^2 \xi\zeta
  + X^2 Y \xi\zeta \cr
  + \frac 14 X^2 Y^2 \xi^2 \zeta^2
	}
	&
\matrix{
1 + 2 X + \hf X^2
  + Y \cr
  + 2 \xi XY
  +
\hf \xi^2 X^2 Y
}
	&
1 + X + Y + \xi XY
\\[10mm]
\t_1    &
\matrix{
1 + \hf X^2 + \hf Y^2 \cr
  + 2 XY (-\xi + \zeta)
  + \frac 14 X^2 Y^2 \xi^2 \zeta^2
}
	&
\matrix{
1 + \hf X^2
  - Y \cr
  -
\hf \xi^2 X^2 Y
}
	&
1 - X - Y + \xi XY
\\[10mm]
\t_2    &
\matrix{
1 - 2 X + \hf X^2
  - 2 Y + \hf Y^2 \cr
  + 2 XY (\xi + \zeta)
  - X Y^2 \xi\zeta
  - X^2 Y \xi\zeta \cr
  + \frac 14 X^2 Y^2 \xi^2 \zeta^2
}
	&
\matrix{
1 - 2 X + \hf X^2
  + Y \cr
  - 2 \xi XY
  + \hf \xi^2 X^2 Y
}
	&
1 + X + Y + \xi XY
\\[10mm]
X
	&
q \,e^{ \sqrt 2 \mass ( x \cosh \eta + t \sinh \eta )}
	&
q \,e^{ \sqrt 2\mass ( x \cosh \eta + t \sinh \eta )}
	&
q \,e^{ 2\mass ( x \cosh \eta + t \sinh \eta )}
\\[1mm]
Y
	&
q' e^{ \sqrt 2\mass ( x \cosh \eta' + t \sinh \eta' )}
	&
q' e^{ 2\mass ( x \cosh \eta' + t \sinh \eta' )}
	&
q' e^{ 2\mass ( x \cosh \eta' + t \sinh \eta' )}
\\[1mm]
\xi
	&
\frac{ (e^\eta - e^{\eta'})^2}{ e^{2\eta} + e^{2\eta'} }
	&
\frac{ \sqrt 2 e^{\eta} + (1+i) e^{\eta'} }
     { \sqrt 2 e^{\eta} - (1+i) e^{\eta'} }
\frac{ \sqrt 2 e^{\eta} + (i-1) e^{\eta'} }
     { \sqrt 2 e^{\eta} - (i-1) e^{\eta'} }
	&
\left(
\frac{ e^{\eta} - e^{\eta'} }{ e^{\eta} + e^{\eta'} }
\right)^2
\\[1mm]
\zeta
	&
\frac{ e^{2\eta} + e^{2\eta'} }{ (e^\eta + e^{\eta'})^2}
	&
	&
\end{array}
\]
\caption{Two soliton solutions}
\label{tab.2sol}
\end{table}

The spectrum of bounded states in a soliton background comprises two
parts, normalisable bound states and unnormalisable scattering
states. It is straightforward to find these from the two soliton
solutions.
We give details here for the states which correspond to type (i)
particles moving in a type II soliton background.

{}From table 1 we take the solution for 2 solitons, one of type I and
one of type II. Since we are considering the type I soliton as a
perturbation of type (i) particles in a background of a stationary
type II soliton, we take $\eta'=0$ and consider the linearised field
$\delta\phi(x,t)$ defined by
\eq
 \phi(x,t) = \phi_{II}(x) + q \delta\phi(x,t) + O(q^2)
\en
We obtain
\eq
\delta\phi(x,t) =
2 (\alpha_0^\vee - \alpha_2^\vee)
e^{\sqrt{2}\mass (x \cosh\eta + t \sinh\eta)}
\frac   { 1 + \frac{\sqrt2\cosh\eta - 1}{\sqrt2\cosh \eta +1} e^{2\mass x} }
	{1 + e^{2\mass x}}
\label{eq.dp}
\en
This solution can correspond to  two different kinds of bounded
solutions.


For $\sqrt2\cosh\eta=1$ the solution is normalisable and corresponds
to a type (i) particle bound to the type II soliton.
The energy of this state is $\mass $.


For $\sqrt2 \mass \cosh\eta = ik$, the solution (\ref{eq.dp})
corresponds to a scattering state $\delta\phi(x,t)$
of a type (i) particle off a type II soliton. We find that
asymptotically
\eq
\begin{array}{rcll}
\delta\phi(x,t) &\sim
	& 2 (\alpha_0^\vee - \alpha_2^\vee)e^{ikx -iwt}
	& x \to -\infty \\[1mm]
\delta\phi(x,t) &\sim
	& 2 (\alpha_0^\vee - \alpha_2^\vee)e^{ikx -iwt + i\delta}
	& x \to +\infty
\end{array}
\en
where the phase shift $\delta$
acquired by the particle as it traverses the
soliton, and the energy $\omega$ of the particle, are given by
\eq
e^{i\delta} =
\frac{ik - \mass }{ik + \mass }
\;,\;\;\qquad\qquad
\omega = \sqrt{ k^2 + 2 \mass ^2 }
\en

It is straightforward to perform this analysis for the other solitons
and particle states.
We summarise the results for both particles in both soliton
backgrounds in tables 2 and 3 below.

\vspace{-\abovedisplayskip}
\begin{tabular}{cc}
\begin{minipage}[t]{6cm}
\[
\begin{array}{c|cc}
e^{i\delta}     & \hbox{type I}         & \hbox{type II}
\\[1mm]
\hline\\[-3mm]
\hbox{type (i)} & \frac{ ik - \sqrt 2\mass }{ik + \sqrt 2\mass }
		& \frac{ ik -  \mass }{ik +  \mass }
\\[1mm]
\hbox{type (ii)} & \left(\frac{ ik - \sqrt 2\mass }{ik + \sqrt 2\mass
}\right)^2
		& \frac{ ik - 2\mass }{ik + 2\mass }
\end{array}
\]
\end{minipage}
&
\begin{minipage}[t]{6cm}
\[
\begin{array}{c|cc}
\omega  & \hbox{type I}         & \hbox{type II}
\\[1mm]
\hline\\[-3mm]
\hbox{type (i)} & 0
		& \mass
\\[1mm]
\hbox{type (ii)} & \sqrt 2\mass
		& 0
\end{array}
\]
\end{minipage}
\\[3cm]
{Table 2: Phase shift $e^{i\delta}$}
&
{Table 3: Energy $\omega$ of normalisable states}
\end{tabular}
\vskip 5mm
Since we take different sets of oscillators in soliton backgrounds to
those in the vacuum, the contribution (\ref{eq.e0}) will differ.
If we choose
\choose
{%
$\delta\phi$ periodic
}{%
$\delta\phi(0,t)=\delta\phi(L,t)$ at the edges of a box of side $L$
}
we have  (for $L$ large and the soliton in the middle of the box)
\choose
{%
$\delta\phi = \delta\phi_k - \delta\phi_{-k}$
}{%
$\delta\phi = \delta\phi_k $
}
with the condition on $k$ that
\eq
kL + \delta(k) = 2 n \pi
\choose
{%
\;,\;\; k \geq 0
}{}
\en
The standard \cite{DHNe1,Raja1} way to compare the zero-point energy
$E_0$
for particles of type $a$ in a soliton background is
to take
\eqq
E_0^{(soliton),a} - E_0^{(vac),a}
&=&
L \int_{-\infty}^\infty \frac{d\!k}{4\pi}
\left(
\sqrt{ k^2 + m_a^2 } - \sqrt{ (k + \delta^{}(k)/L)^2 + m_a^2}
\right)
\nonumber\\
&=&
- \int_{-\infty}^\infty \frac{d\!k}{4\pi}
\frac{k \delta^{}(k)}{\sqrt{ k^2 + m_a^2} }
+ O(\frac1L)
\nonumber\\
&=&
\left[
	-\frac{1}{4\pi}\delta^{}(k) \sqrt{ k^2 + m_a^2}
\right]_{-\infty}^\infty
+ \int_{-\infty}^\infty \frac{d\!k}{4\pi} {\sqrt{ k^2 + m_a^2} }
	\frac{d}{dk} \delta^{}(k)~~~~~~~~~~~
\label{eq.edif}
\enn
These are easy to evaluate for the phase shifts we have found,
which are all of the form
\eq
\exp(\, i\delta(k) \,) =
\prod_j \frac{i k - a_j }{i k + a_j }
\en
for which
\eq
\left[
	\delta^{}(k) \sqrt{ k^2 + m_a^2}
\right]_{-\infty}^\infty
=
 \lim_{k \to \infty}  k\delta^{}(k)
+ \lim_{k \to -\infty} k\delta^{}(k)
=
2 \sum_j a_j
\label{eq.ps2}
\en
and
\eq
\frac{d}{dk} \delta(k) = - 2 \sum_j \frac{a_j}{k^2 + a_j^2}
\en
The  term in (\ref{eq.edif}) is no longer quadratically
divergent, but is still
logarithmically divergent.
We still need to consider the effects of the divergence in the
potential, but  these  will  turn  out  to have no effect on the soliton mass
ratios.

\subsection{Normal ordering}

After subtracting the zero-point energy from the Hamiltonian we still
have a divergent theory.
In the vacuum sector we have the relation \cite{Cole1}
\eq
\mno
\exp( \gamma\cdd\phi )
\mno
=
\exp\left( \,-\hf \cev{vac}
	\;\gamma\cdd\phi\, \gamma\cdd\phi\;
		\vec{vac} \,\right)
\,
\exp(\gamma\cdd\phi)
\en
With our boundary conditions $\delta\phi$ periodic
we have
\eq
\cev{vac} \;\xi_a\cdd\phi\, \xi_a\cdd\phi \; \vec{vac}
=
\Delta_a
=
\int_{-\Lambda}^\Lambda
\frac{d\!k}{4\pi}
\frac 1{\sqrt{ k^2 + m_a^2}}
\en
where $\xi_a$ are the unit eigenvectors of the mass matrix.
To zeroth order in $\b^2$ this gives a correction to the soliton mass as
\eq
\int \! d\!x
\sum_{a=0} n_a
\, \left( \,
\Big( \exp( \b \alpha_a\cdd\phi_{soliton}) - 1 \Big)
\sum_{b=1}
\hf \mass^2 (\xi_b \cdd \alpha_a)^2 \Delta_b
\, \right)
\label{eq.noc}
\en
The calculation of this is simplified since, as in the $A_n$ case
\cite{Holl2}, the integrals in
\reff{eq.noc} are independent of the root $\alpha_a$
for stationary soliton solutions,
as can  be  easily seen from the equations of motion.  Using
\cite{OTUn2} for the soliton mass $M_j$
we find the universal result
\eq
\int dx ( \exp(\beta\alpha_a\cdd\phi_{soliton}) -1 )
=
\frac{ \beta^2}{ 2 \mass^2 h } M_j
\en
where $h$ is the Coxeter number.  Hence the normal ordering contributions for
each  soliton are proportional  to  the  soliton  masses  and  cancel  in  any
calculation of the mass ratios.

\section{Adding the contributions}

We can now add the various contributions and check that the answer is
finite. We list in table 4 the various contributions and their sums for the
two solitons
\[
\begin{array}{r|cc}
\parbox{4cm}{Contributions from}
	& \hbox{type I}
	& \hbox{type II} \\[1mm]
\hline\\[-3mm]
\parbox{4cm}{Bound states\\ of type (i) particles}
	&       & \frac{\mass}2 \\[.7cm]
\parbox{4cm}{Bound states\\[-1mm] of type (ii) particles}
	& \frac{\mass}{\sqrt 2} &       \\[.7cm]
\parbox{4cm}{Scattering states\\[-1mm] of type (i) particles}
	& -\frac{1}{4\pi} 2\sqrt 2\mass
	& -\frac{1}{4\pi} 2\mass
	\\
	& - \int \!\frac{dk}{4\pi}
		\frac{ 2\sqrt 2\mass \sqrt{ k^2 + 2\mass^2}}
		{k^2 + 2\mass^2}
	& - \int \!\frac{dk}{4\pi}
		\frac{ 2 \mass \sqrt{ k^2 + 2\mass^2}}
		{k^2 + \mass^2}
	\\[.7cm]
\parbox{4cm}{Scattering states\\[-1mm] of type (ii) particles}
	& -\frac{1}{4\pi} 4\sqrt 2\mass
	& -\frac{1}{4\pi} 4\mass
	\\
	& - \int \!\frac{dk}{4\pi}
		\frac{ 4\sqrt 2\mass \sqrt{ k^2 + 2\mass^2}}
		{k^2 + 2\mass^2}
	& - \int \!\frac{dk}{4\pi}
		\frac{ 4 \mass \sqrt{ k^2 + 4\mass^2}}
		{k^2 + 4\mass^2}
	\\
\end{array}
\]
As can be seen when all soliton mass ratio is calculated
and the integrals combined appropriately, the
divergent terms cancel and the total answer is finite. In all we find
\eq
\frac{M_I}{M_{II}}
 \,=\,
\sqrt 2\left( 1 + \frac{\b^2}{64} \right)
\en
As can be seen the ratios of the soliton masses does not remain
constant when 1 loop quantum corrections are included.
It is a  remarkable  coincidence that this change in the
mass ratio is equal to that of the
massive particles in the theory.

\section{Comments and Conclusions}

If correct, the observation that the soliton mass raio does no remain
constant introduces some problems in the construction of soliton S
matrices. The R-matrix approach in which all the `group' information is
encoded in an R-matrix seems unlikely to suceed as this method relies
on the ratios of the soliton masses having theor classical values.

It is possible that there are some unforeseen difficulties in the
application of this method to the non-simply-laced Affine Toda
theories which will return the masses to their classical ratio.
Alternatively the soliton scattering is much more complicated than
hitherto guessed.

\leftline{\bf Acknowledgements}

I would like to thank
N.J.\ Burroughs,
P.\ Johnson,
N.J.\ MacKay
and
D.I.\ Olive
for helpful conversations, comments and criticism at various stages.
This work was supported by a fellowship from St.\ John's College,
Cambridge.
I would especially like to thank everyone at
IAFE, University of Buenos Aires, for their very
kind hospitality while this work was completed.


\begin{thebibliography}{10}

\bibitem{Kac1}
V.~Kac,
\newblock Infinite dimensional Lie algebras,
\newblock Cambridge University Press, 1985.

\bibitem{AfTod}
H.~Braden, E.~Corrigan, P.~Dorey and R.~Sasaki,
Phys.\ Lett.\ B227 (1989) 411;
Nucl.\ Phys.\ B288 (1990) 689;
Nucl.\ Phys.\ B356 (1991) 469;
P.~Dorey,
Nucl.\ Phys.\ B358 (1991) 654;
Phys.\ Lett.\ B312 (1993) 291;
G.W.~Delius, M.T.~Grisaru and D.~Zanon,
Phys.\ Lett.\ B277 (1992) 414;
Nucl.\ Phys.\ B382 (1992) 365.


\bibitem{Cole1}
S.~Coleman,
\newblock Phys. Rev. D11 No. 3  (1975) 2088.


\bibitem{Holl3}
T.~J. Hollowood,
\newblock Nucl. Phys. B384 (1992) 523.

\bibitem{MMcg1}
N.~J. MacKay and W.~A. McGhee,
\newblock Int. J. Mod. Phys. A8 (1993) 2791,
\newblock Erratum Int.\ J.\ Mod.\ Phys.\ A8 (1993) 3830; hep-th/9208057.

\bibitem{CZhu1}
D.~G. Caldi and Z.~Zhu{\it,
\newblock Multi--soliton solutions of affine Toda models},
\newblock SUNY, Buffalo Preprint UB-TH-0193 (1993),
\newblock hep-th/9307175.

\bibitem{CFGo1}
C.~P. Constantinidis et.\ al.\ 
\newblock Phys. Lett. B298 (1993) 88,
\newblock hep-th/9207061.

\bibitem{ACFe1}
H.~Aratyn et.\ al.\ 
{\it,
\newblock Construction of affine and conformal affine Toda solitons by Hirota's
  method},
\newblock U. Chicago Preprint IFT-P-020-93, UICHEP-TH-93-3, C93-01-10 (1993),
\newblock Presented at 7th Summer School Jorge Andre Swieca: Particles and
  Fields, Sa\~o Paulo, Brazil, 10-23 Jan 1993; hep-th/9304080.

\bibitem{ACFe2}
H.~Aratyn et.\ al.\ 
\newblock Nucl. Phys. B406 (1993) 727,
\newblock hep-th/9212086.

\bibitem{OTUn1}
D.~I. Olive, N.~Turok and J.~W.~R. Underwood{\it,
\newblock Affine Toda solitons and vertex operators},
\newblock University College, Swansea, Preprint PUP-TH-93-1392 (1993).

\bibitem{Holl2}
T.~J. Hollowood,
\newblock Phys. Lett. B300 (1993) 73,
\newblock hep-th/9209024.

\bibitem{DHNe1}
R.F.~Dashen, B.~Hasslacher and A.~Neveu,
\newblock Phys.\ Rev.\ D10 (1974) 4114;
\newblock Phys.\ Rev.\ D11 (1975) 3424;
\newblock Phys.\ Rev.\ D12 (1975) 2443.

\bibitem{Raja1}
R.~Rajaraman,
\newblock Solitons and Instantons: An introduction to solitons and instantons
  in quantum field theory,
\newblock North--Holland, 1982.

\bibitem{OTUn2}
D.~I. Olive, N.~Turok and J.~W.~R. Underwood,
\newblock Nucl. Phys. B401 (1993) 663.

\end{thebibliography}

\end{document}